# A SIMPLE TRANSITION-FREE LATTICE OF AN 8 GEV PROTON SYNCHROTRON*

W. Chou[#], Fermilab, Batavia, IL 60510, U.S.A.

*Abstract*

A transition-free lattice is a basic requirement of a high-intensity medium-energy (several GeV) proton synchrotron in order to eliminate beam losses during transition crossing. An 8 GeV synchrotron is proposed as a principal component in an alternative hybrid design of Project-X [1]. This machine would be housed in the Fermilab antiproton source enclosure replacing the present Debuncher. A simple doublet lattice with high transition gamma has been designed. It uses just one type of dipoles and one type of quadrupoles (QF and QD are of the same length). It has no transition crossing. It has a triangular shape with three zero dispersion straight sections, which can be used for injection, extraction, RF and collimators. The beta-functions and dispersion are low. This lattice has plenty of free space for correctors and diagnostic devices, as well as good optical properties including large dynamic aperture, weak dependence of lattice functions on amplitude and momentum deviation.

## INTRODUCTION

The Fermilab Tevatron Collider experiments will be completed in 2010 or 2011. Presently Fermilab is pursuing future facilities on both the energy and intensity frontiers. The leading candidate on the intensity frontier is a multi-MW proton source, nicknamed Project-X. There exist several designs. One is an 8 GeV superconducting RF (SRF) H$^-$ linac [2]. Another is an early design that uses an 8 GeV rapid cycling synchrotron (RCS) [3]. Each design has its pros and cons. A third approach is a hybrid design, which consists of a 2 GeV SRF H$^-$ linac and a 2-8 GeV RCS. This approach presents a number of attractive features [1].

In the hybrid design, the RCS would be housed in the Fermilab antiproton source enclosure replacing the present Debuncher. This places several constraints on the RCS lattice design, because it must fit to the Debuncher footprint and has the same circumference (505 m) and same shape (triangular). Besides, it must be transition-free in order to avoid any potential beam loss during transition crossing in high intensity operation. This paper presents a lattice design that meets these requirements and also has good optical properties.

## DESIGN GUIDANCE

The lattice is simple and cannot get any simpler, since only one bend type and one quad type are required. The focusing and defocusing quads are identical. The lattice adopts a missing-magnet modular structure based on the following concept.

Two equal bends separated by 180° in phase constitute an achromatic bend. A module comprised of three 90° phase-advance cells

Module = (cell + bend, cell without bend, cell + bend)

is achromatic (zero dispersion in, zero dispersion out). It also places the bends in regions of low dispersion and puts high dispersion in the cell without bend. A ring can then be constructed with an arbitrary number of such modules and don't require dispersion suppressors. The non-bend cells can contain sextupoles, correctors and collimators (high dispersion). The non-bend cells and modules are automatically matched and can be mixed to taste. In this design, four modules are used in an arc followed by four non-bend cells in a straight for a period. Three periods form a triangular ring, which fits nicely in the Debuncher enclosure.

Elimination of sextupole generated "geometrics" requires odd-$\pi$ separation of sextupole pairs in the horizontal plane. The 3-cell module based on the 180° achromat property automatically has a phase shift of 270°. Therefore the odd-$\pi$ condition is met with 540° or two module separation. In the vertical plane, one can correct geometrics with either odd-$\pi$ or even-$\pi$ pairs. The odd-odd case is a clear cut. The odd-even case also works but will not be elaborated here. Hence the missing-magnet scheme must be based either on a 90°/90° phase advance cell (270°/270° module) or a 90°/60° phase advance cell (270°/180° module). The former is chosen in this design.

Figure 1 is a 2D illustration of the lattice. Figure 2 shows the lattice structure and lattice functions. Table 1 lists the lattice structure, magnets and optical properties.

## LATTICE DESCRIPTION

The main features of this lattice are as follows:
- Triangular shape to fit to the Debuncher footprint (505 m, h = 90)
- No transition crossing ($\gamma_t$ = 18.6)
- Zero dispersion straights
- Simple:
  - 1 type of dipole
  - 1 type of quad (QF and QD of the same length)
  - 48 cells, all cells of the same length
- Doublet lattice, 90° phase advance per cell
- Modular structure:
  - 3 cells per module, missing dipole in mid-cell, 270° phase advance per module
  - 4 modules per arc, 6$\pi$ phase advance per arc
    $\Rightarrow$ No dispersion suppressor

- 4 cells per straight
- Each period with one arc and one straight
- 3 periods for the ring
- Peak bending field 1.5 T, peak gradient 10.26 T/m
- Plenty of free space (for RF, injection/extraction, collimators, sextupoles, correctors, diagnostics)
  - Long: 7.55 m × 24
  - Short: 1.18 m × 48
  - Mini: 0.5 m × 48
- Low beta-functions (16 m) and dispersion (2.4 m)
- Good optical properties (large dynamic aperture, weak dependence of lattice functions on amplitude and dp/p)

## DISCUSSION

The modular design adopted by this lattice is rather flexible. It can be used for lattices of different sizes and various shapes – not only triangular, but also racetrack, rectangular, etc.

It is the simplest lattice one can imagine, because all bending magnets and all quadrupoles are identical. This makes magnet fabrication simpler and cheaper, machine tuning and operation easier.

## ACKNOWLEDGEMENTS

The author would like to thank C. Prior for making a 2D lattice plot in Figure 1.

## REFERENCES

[1] W. Chou, "A Hybrid Design of Project-X," this conference.
[2] S. Holmes, "Project-X," this conference.
[3] "Proton Driver Study II – Part 1," Fermilab-TM-2169. (May 2002)

Table 1: Lattice Structure, Magnets and Optical Properties

| | |
|---|---|
| Circumference | 505.294 m |
| Shape | Triangle |
| Free space | 0.50 m × 48 |
| | 1.17548 m × 48 |
| | 7.55096 m × 24 |
| Dipoles (1 type) | 5.2 m × 24 = 124.8 m |
| Peak field | 1.5 T |
| Full gap | 7.5 cm |
| Quadrupoles (1 type) | 1.238 m × 96 = 118.848 m |
| Peak gradient | 10.26 T/m |
| Sextupole | 24 HS, 24 VS |
| Cell | Doublet |
| | 10.527 m, 90°/90° |
| Arc module | 3-cell, missing D in mid-cell |
| | 31.581 m, 270°/270° |
| Arc | 4 modules |
| | total 12 cells, 126.324 m |
| Straight | 4 doublet cells, 42.108 m |
| Ring | 48 cells |
| $\beta_x/\beta_y$ | 15.6 m / 16.2 m |
| Dispersion | 2.4 m |
| $\gamma_t$ | 18.6 |
| $\xi_x/\xi_y$ | −15.1 / −15.5 |
| $\nu_x/\nu_y$ | 11.98 / 12.21 |
| Dynamic aperture | 350 π |
| $\Delta\nu$ vs. $\Delta p/p$ | < 0.005 at ±1% |
| $\Delta\beta$ vs. $\Delta p/p$ | < 0.4 m at ±1% |
| $\Delta D$ vs. $\Delta p/p$ | < 0.14 m at ±1% |
| $\Delta\gamma_t$ vs. $\Delta p/p$ | < 0.4 at ±1% |
| $\Delta\nu$ vs. amplitude | 0.183      -0.423 |
| | -0.423      0.332 |
| Space charge | $\Delta\nu = 0.15$ at injection |
| 3rd order resonance | Cancelled |

___________________
*Work supported by Fermi Research Alliance, LLC under Contract No. DE-AC02-07CH11359 with the United States Department of Energy.
#chou@fnal.gov

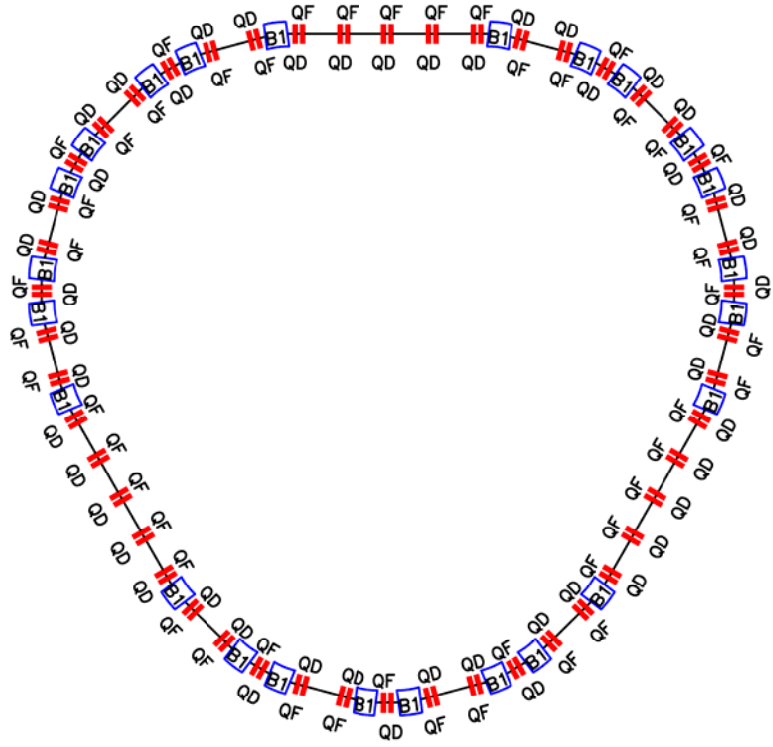

Figure 1: A 2D illustration of the lattice.

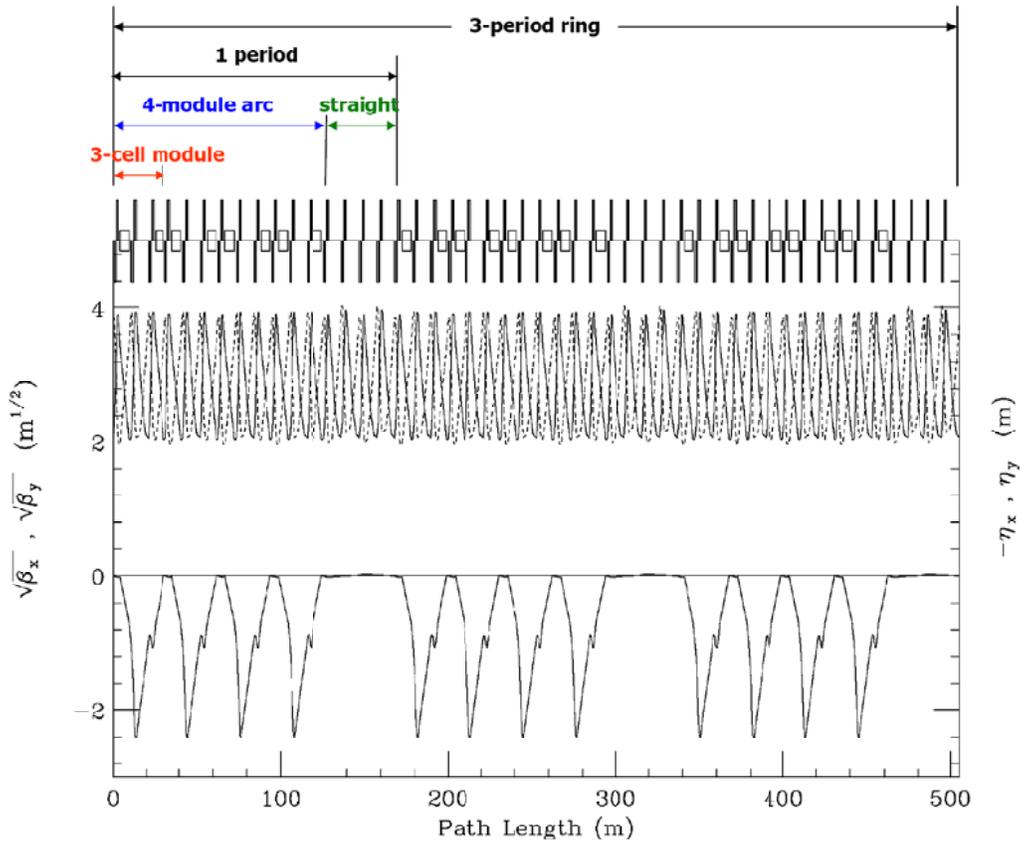

Figure 2: Lattice structure and lattice functions.